\documentstyle[12pt]{article}

\def \scpage{\null}

\begin{document}

\setlength{\baselineskip}{19.4pt}

\rm
\null \vspace{50pt}

\centerline{\Large\rm  Up-Down Quark Mass Difference Effect}
\vspace{8pt}
\centerline{\Large\rm  in Nuclear Many-Body Systems}
\vspace{35pt} \rm

\centerline{
  S. N{\footnotesize AKAMURA}$^a$, K. M{\footnotesize UTO}$^b$,
  M. O{\footnotesize KA}$^b$, S. T{\footnotesize AKEUCHI}$^c$
  and T. O{\footnotesize DA}$^b$}
\vspace{15pt} \rm

\centerline{$^a$ \it Institute for Nuclear Study,
  University of Tokyo, Tokyo 188, Japan}
\centerline{$^b$ \it Department of Physics,
  Tokyo Institute of Technology, Tokyo 152, Japan}
\centerline{$^c$ \it Department of Public Health
  and Environmental Science, }
\centerline{\it Tokyo Medical and Dental University, Tokyo 113, Japan}

\null \vfill \null

$$\vbox{
\hsize 13.0truecm
\centerline{\bf Abstract }
\rm
\noindent
A charge-symmetry-breaking nucleon-nucleon force
due to the up-down quark mass
difference is evaluated in the quark cluster model.
It is applied to the shell-model calculation
for the isovector mass shifts of isospin multiplets
and the isospin-mixing matrix elements in 1s0d-shell nuclei.
We find that the contribution of the quark mass difference effect
is large and agrees with experiment.
This contribution may explain the Okamoto-Nolen-Schiffer
anomaly, alternatively to the meson-mixing contribution,
which is recently predicted to be reduced by the large
off-shell correction.
\vspace{15pt}
}$$

\vspace{80pt}
\null

\clearpage

\rm
Charge-symmetry-breaking (CSB) terms is required
in the nuclear force for explaining several phenomena,
e.g., the difference between the proton-proton and neutron-neutron
scattering lengths \cite{111Mill90} and the anomaly of mass differences
of several mirror nuclei, called Okamoto-Nolen-Schiffer (ONS)
anomaly \cite{117Okam64,218Nole69,115Sato76,123Shlo78}.
These experimental data imply that the nuclear force between
two neutrons is slightly more attractive than between two protons.
\par\rm
Three of the present authors
have made an extensive analysis of isovector mass shifts
and isospin-mixing matrix elements in 1s0d-shell nuclei.
It was shown that experimental values of these quantities
are well explained \cite{inc3} by
a short-range CSB force, but not by a long-range force.
According to these findings, we
look for the origin of a short-range CSB force in this report.
\par\rm
The experimental isovector mass shift, $b(\nu,T)$, is calculated by
letting the isospin multiplet mass equation,
\begin{displaymath}
E(\nu,T,T_z)=a(\nu,T)+b(\nu,T)T_z+c(\nu,T)T_z^2,
\end{displaymath}
reproduce the masses, $E(\nu,T,T_z)$, of the $2T+1$ members of
the nuclear isospin multiplet.
In the above equation, $\nu$ represents the quantum numbers
other than the isospin, $T$, and its third component, $T_z$.
\par\rm
Several well-known CSB sources have contributions
to the isovector mass shifts.
Before considering the nuclear-CSB force, we subtract
the contributions of the electromagnetic interactions (EM)
and of the isovector single particle energy (ISPE).
They are calculated by the analysis of
1s0d-space shell model \cite{inc3,Dthesis}.
The EM contribution is evaluated in taking account of the Coulomb force
between the protons with the charge form factor correction,
the electromagnetic spin-orbit force and the magnetic spin-spin
contact interaction with the magnetic form factor of the nucleon.
The explicit formulae are given in ref.\cite{inc3}.
The ISPE represents the CSB interactions between
the valence nucleon and the $^{16}$O-core.
Subtracting them, we obtain the ``experimental" data, shown in Fig.\ 1,
which is to be compared with the nuclear-CSB contribution.
\par\rm
We find a systematic behaviour, called ``zigzag behaviour",
i.e., that the ``experimental" values of the isodoublets are reduced
for $A=4n+1$ while for $A=4n+3$ they are enhanced.
It is demonstrated in the previous studies \cite{inc3,Dthesis}
that a short-range CSB force is the most probable source of
the ``zigzag behaviour".
\par\rm
A short-range nuclear-CSB force
can be provided by the exchange of the mixed $\rho$-$\omega$ complex.
It is extremely short-ranged, and explains at least half of the
``zigzag behaviour" \cite{Dthesis}.
The meson-mixing potential is also used to explain other
phenomena, e.g., the ONS anomaly
\cite{86Blun87,85Wu90,82Suzu92}.
Goldman {\it et al.} \cite{302Gold92} argued, however,
that an off-shell correction
reduces the meson-mixing amplitude by a large factor.
The correction is so large as to kill the meson-mixing contribution.
Other calculations \cite{368Piek93,369Krei93,367Hats94} also verify
such an off-shell effect.
It is, however, still controversial, because another analysis
indicates the strong off-shell effect being inconsistent with
the observed $q^2$-dependence of $\rho$-$\gamma^*$ coupling \cite{Mill94}.
Further studies seem to be necessary.
\par\rm
In the present article, we assume that the meson-mixing
contribution is negligible, and examine
another short-range CSB force due to the up-down quark
mass difference, called quark effect (QE).
Precisely speaking, it is a contribution of the quark mass difference
in the gluon-exchange interaction between the valence quarks,
while the meson-mixing may contain effects of the quark mass
difference as well \cite{111Mill90}.
Such a quark CSB force has been used to explain
the difference between the proton-proton and neutron-neutron
scattering lengths \cite{279Chem84} and the
ONS anomaly \cite{Koch85}.
In this study, we apply the quark CSB force to the shell model
calculation for the mass shifts of nuclear isospin multiplets.
\par

\vspace{20pt}

\rm
The quark CSB potential is calculated in the nonrelativistic
potential quark model of baryons \cite{279Chem84}.
The model consists of the standard interaction hamiltonian,
which contains a quark confining potential as well as
one-gluon exchange interaction.
The confining potential is assumed to be independent of isospin,
i.e., flavour of the quarks \cite{111Mill90,279Chem84}.
The one-gluon exchange interaction contains an isospin-symmetry-breaking
term in the hyperfine contact interaction,
\begin{equation}
H^{({\rm q}_i {\rm q}_j)}_{\rm HC} = -(\lambda_i \cdot \lambda_j)
   \frac{\pi\alpha_{\rm s}}{6 m_i m_j}
   (\vec{\sigma}_i \cdot \vec{\sigma}_j) \delta^{(3)}(r_{ij}) ,
\label{eq:hypcnct}
\end{equation}
where $\lambda_i$, $\vec{\sigma}_i$ and $m_i$ are
the colour SU(3) generator, Pauli spin matrix
and mass of the constituent quark, q$_i$, and
$\alpha_{\rm s}$ is the strong coupling constant.
Other terms are estimated to have negligible contributions to
the isospin-dependent nucleon-nucleon (NN) interaction \cite{279Chem84}.
This contact interaction yields a short-range NN
force which has the range of the
nucleon size.
\par\rm
The two nucleon system is represented by a quark-cluster
wave function composed by two three-quark clusters \cite{279Chem84}.
The internal wave function of each nucleon is approximated by
a Gaussian, and the internal variables are integrated out
to obtain the NN potential
from the quark-quark interaction.
The obtained potential depends on the NN relative
coordinate as well as the spin and isospin quantum numbers.
\par\rm
It is found in the calculation of Chemtob and Yang
(see Fig.\ 2 of ref.\cite{279Chem84}) that the local term
of the hyperfine contact interaction is the leading term and
represents the whole CSB interaction approximately.
In this article, therefore, we deal only with the local term of
the hyperfine contact interaction, and the other terms are neglected.
\par\rm
In order to apply the potential to the shell-model calculation,
we have to prepare the potentials for
higher partial waves than the S-states.
They are numerically calculated \cite{Take89},
and we obtain a quark CSB potential,
\begin{equation}
v^{\rm (N_1 N_2)}_{\rm QE} =
   \sqrt{\frac{3}{\pi}} \frac{\beta^3 \alpha_{\rm s}}{\hat{m}^2}
   \, {\rm exp}(-\frac{3}{4} \beta^2 r_{12}^2)
   \, \frac{1}{4}[\tau_z(1)+\tau_z(2)] \frac{\delta m}{\hat{m}}
      [1-\frac{5}{27} \vec{\sigma}(1)\cdot\vec{\sigma}(2)] ,
\label{eq:vQE(o)}
\end{equation}
where $\beta^{-1}$ is the nucleon size parameter of
the three-quark cluster, and $r_{12}$ is the distance
between the centers of the two clusters,
$\tau_z(i)$ is the third component of the Pauli matrix for
the nucleon isospin, and $\vec{\sigma}(i)$ is for the nucleon spin.
$\delta m = m_{\rm d} - m_{\rm u}$ is the up-down quark
mass difference and $\hat{m} = \frac{1}{2} (m_{\rm d} + m_{\rm u})$
is the average of the masses.
The input parameters are taken from ref.\cite{279Chem84},
$\delta m = 6$ MeV, $\hat{m} = 330$ MeV, $\alpha_{\rm s} = 1.624$
and $\beta^{-1} = 0.616$ fm.
\par\rm
The nuclear matrix element of the QE potential gives
the isovector mass shift,
\begin{displaymath}
b^{\rm QE}(\nu,T) = \frac{1}{\sqrt{(2T+1)T(T+1)}}
                    \langle \nu,T|| v_{\rm QE} ||\nu,T \rangle ,
\end{displaymath}
in the first order perturbation, where the matrix element
$\langle \nu,T|| v_{\rm QE} ||\nu,T \rangle$ is
reduced with respect to the isospin.
\par\rm
The nuclear wave function, $|\nu,T \rangle$, is calculated
\cite{inc3,Dthesis} with Wildenthal's effective hamiltonian
\cite{37Wild84} in the complete 1s0d-shell space.
Because the short-range QE potential is integrated,
the calculation is sensitive to the short-range structure of
the relative wave function of the two nucleons.
We include a short-range correlation by a correlation
function of ref.\cite{155Mill76} multiplied to
the relative two-nucleon wave function.
\par\rm

\vspace{20pt}

\rm
The calculated QE contributions are shown in Fig.\ 2.
The QE contributions are around 100 keV,
and the average ratio of QE/Coulomb contributions for
143 multiplets is 5.3\%.
This result is consistent with the expected contribution
which is introduced phenomenologically to explain
the ONS anomaly in literatures
\cite{115Sato76,123Shlo78}, and consistent with the previous
calculation of the quark effect for the anomaly \cite{Koch85}.
\par\rm
The ``zigzag behaviour" seen in the ``experimental" mass shifts
(Fig.\ 1) is also reproduced in the calculation (Fig.\ 2).
Namely, the quark contributions are larger for the $A=4n+3$
isodoublets, and smaller for $A=4n+1$.
This behaviour is due to the short-range nature of the
quark CSB force \cite{inc3}.
\par\rm
We need to comment on the calculation of the ISPE contribution.
The ISPE used in Fig.\ 1 is determined by a $\chi^2$-fitting to the
experimental mass shifts, including the QE contribution
by the prescription of ref.\cite{Dthesis}.
By this procedure, the ISPE could be adjusted to produce
the ``zigzag behaviour" given by the QE contribution,
even if the original data would not have a ``zigzag" feature.
To examine this, we have made another $\chi^2$-fitting
without any nuclear-CSB contribution,
and still found the ``experimental zigzag behaviour"
(see ref.\cite{inc3,Dthesis}).
\par

\vspace{20pt}

\rm
The quark CSB potential of eq.(\ref{eq:vQE(o)}) is also
applied to off-diagonal matrix elements, i.e.,
isospin-mixing matrix elements.
The results are given in Table 1, with experimental
values extracted from the strength of the isospin-forbidden beta decays.
We find that the calculated matrix elements agree
with most of the experimental values.
\par\rm
{}To the off-diagonal matrix elements, the quark contribution
is found to be large, comparable with
the electromagnetic contributions (EM) in the table,
which are roughly equal to the contributions of the Coulomb force.
By contrast, the mass shifts are dominated by the Coulomb contributions.
Such large contributions to the off-diagonal matrix elements
can be attributed to the short-range feature of the quark CSB force
\cite{inc2,inc3}.
\par\rm
In particular, the beta decay of $^{24}$Al (Table 1)
indicates a large isospin-mixing in an excited state
of the daughter nucleus $^{24}$Mg.
Our calculation also gives a large QE contribution to this
matrix element, and thus is consistent with experiment.
\par

\vspace{20pt}

\rm
In summary, the isovector mass shifts and isospin-mixing
matrix elements relevant to the isospin-forbidden beta decays
are calculated considering the effect of the up-down quark
mass difference in the direct gluon-exchange process.
The charge-symmetry-breaking (CSB) nucleon-nucleon potential
due to the quark mass difference is constructed
with the quark cluster model, and is applied to the shell model
calculation of the 1s0d-shell space.
\par\rm
The quark contributions are found to be about 5\% of the
Coulomb contributions to the isovector mass shifts of
the nuclear isospin multiplets.
It is consistent with the expected contributions
\cite{115Sato76,123Shlo78} to explain
the anomaly of mass differences of the mirror nuclei,
known as the Okamoto-Nolen-Schiffer anomaly.
Also, the quark contribution is found to have a systematic behaviour
in the contributions to the mass shifts of the isospin doublets,
i.e., for $A=4n+1$ the mass shifts are reduced while
for $A=4n+3$ they are enhanced.
This behaviour is consistent with the experimental one.
\par\rm
The calculated values of the isospin-mixing matrix elements
are consistent with the most of the experimental values
extracted from the isospin-forbidden beta decays.
The quark CSB force is found to have large contributions
to the isospin-mixing matrix elements, comparable with
the Coulomb contributions.
A particularly large value of the experimental matrix element
of $^{24}$Mg is found to be explained by the large quark contribution.
These findings strongly suggest the existence of a short-range
CSB interaction.
\par\rm
The quark mass difference effect is, of course, not a unique
possible source of the short-range CSB force.
However, it seems to have large contribution to the observables,
and have a favourable feature for explaining the known data.
\par

\vspace{20pt}

\rm
The numerical calculations were performed with the FACOM M780
computer system at the Institute for Nuclear Study, University of Tokyo.
This study is partly supported by a Grant-in-Aid for Scientific
Research (05243204 and 06234206) from the Ministry of Education, Science
and Culture (Monbusho), and one of the authors (S.N.) is
supported by a scholarship from the Soryushi Shogakukai.
\par

\vskip 30pt
\scpage

\vfil\eject

{\baselineskip 17pt
\noindent
Table 1 \ Isospin-mixing matrix elements in keV and their
decompositions into the contributions of quark effect (QE),
electromagnetic interactions (EM) and others.
The EM consists of a large Coulomb contribution
plus the other small contributions, and contains the
isotensor terms of the electromagnetic interactions.
``Other" consists of the ISPE contribution and the isotensor
contribution due to the pion mass difference \cite{incm1}.
Experimental values are extracted from the strengths
of isospin-forbidden beta decays
of the given initial states \cite{30Perl78}-\cite{69Albu79}.
\par
\begin{center}
\begin{tabular}{clrrrrrrrc}
\hline
& initial state
  & $^{19}$O  & $^{20}$F \quad  & $^{24}$Al$^{\rm m}$
  & $^{24}$Na & $^{24}$Al & $^{27}$Mg & $^{28}$Mg & \\
\hline
& QE    & $-12.2$ & $ 11.5$ & $ 7.4$ & $ 5.5$ & $ 36.8$ & $  2.6$ & $ -8.2$ \\
& EM    & $  3.3$ & $  8.3$ & $10.0$ & $ 6.2$ & $ 43.3$ & $-16.0$ & $-19.2$ \\
& other & $ 25.5$ & $-11.6$ & $-8.3$ & $-3.3$ & $-21.6$ & $ 41.1$ & $ 67.2$ \\
& total & $ 16.6$ & $  8.2$ & $ 9.1$ & $ 8.5$ & $ 58.5$ & $ 27.8$ & $ 39.7$ \\
& experiment & $20(10)$ & $14^{+29}_{-14}$ & $49(5)$
  & $5.4(22)$ & $106(40)$ & $3.6^{+57}_{-3.6}$ & $20.6(16)$ & \\
\hline
\end{tabular}
\end{center}
\par
}

\vskip 120pt minus 100pt
\scpage

\noindent
{\bf Figure captions:}
\vspace{15pt}
\\

\vbox{
\hsize 10.3truecm
\baselineskip 18pt

\rm
\noindent
Fig.\ 1 \ The experimental isovector mass shifts
\cite{173Waps85a174Waps88,175Ajze87,176Endt78a164Endt90}
after subtracting the electromagnetic contributions and the isovector
single-particle-energy contribution in keV.
The isodoublets ($T=1/2$) are denoted by the filled circles, and
the others ($T>1/2$) are by the crosses.
\par

\vspace{20pt}

\rm
\noindent
Fig.\ 2 \ The quark contribution to the isovector mass shifts in keV.
Notations are the same as in Fig.\ 1.
\par

}

\end{document}